\documentclass[table,twocolumn]{aastex62}
\usepackage{amstext}
\usepackage{amsmath}
\usepackage{apjfonts}
\usepackage[capbesideposition={top}]{floatrow}
\usepackage{float}
\usepackage{cancel}
\usepackage{floatrow}
\newcommand{\secref}[1]{\S\ref{#1}}

\newcommand{\beqar}{\begin{eqnarray}}
\newcommand{\eeqar}{\end{eqnarray}}

\newcommand{\rg}{r_{\rm g}}

\newcommand{\pg}{p_{\rm g}}

\newcommand{\alphaeff}{\alpha_{\rm eff}}
\newcommand{\alpham}{\alpha_{\rm M}}

\newcommand{\beq}{\begin{equation}}
\newcommand{\eeq}{\end{equation}}

\definecolor{nick}{HTML}{006400}

\begin{document}
\title{Nozzle Shocks, Disk Tearing and Streamers Drive Rapid Accretion in 3D GRMHD Simulations of Warped Thin Disks }

\correspondingauthor{Nick Kaaz}
\email{nkaaz@u.northwestern.edu}

\author[0000-0002-5375-8232]{Nicholas Kaaz}
\affiliation{Department of Physics \& Astronomy, Northwestern University, Evanston, IL 60202, USA}
\affiliation{Center for Interdisciplinary Exploration \& Research in Astrophysics (CIERA), Evanston, IL 60202, USA}

\author{Matthew T.P. Liska}
\affiliation{Institute for Theory and Computation, Harvard University, 60 Garden Street, Cambridge, MA 02138, USA}

\author[0000-0003-2982-0005]{Jonatan Jacquemin-Ide}
\affiliation{Center for Interdisciplinary Exploration \& Research in Astrophysics (CIERA), Evanston, IL 60202, USA}

\author[0000-0001-5064-1269]{Zachary L. Andalman}
\affiliation{Center for Interdisciplinary Exploration \& Research in Astrophysics (CIERA), Evanston, IL 60202, USA}
\affiliation{Department of Physics, Yale University, New Haven, CT 06520, USA}

\author{Gibwa Musoke}
\affiliation{Anton Pannekoek Institute for Astronomy, University of Amsterdam, Science Park 904, 1098 XH Amsterdam, The Netherlands}

\author[0000-0002-9182-2047]{Alexander Tchekhovskoy}
\affiliation{Department of Physics \& Astronomy, Northwestern University, Evanston, IL 60202, USA}
\affiliation{Center for Interdisciplinary Exploration \& Research in Astrophysics (CIERA), Evanston, IL 60202, USA}

\author[0000-0002-4584-2557]{Oliver Porth}
\affiliation{Anton Pannekoek Institute for Astronomy, University of Amsterdam, Science Park 904, 1098 XH Amsterdam, The Netherlands}        

\begin{abstract} 
The angular momentum of gas feeding a black hole (BH) may be misaligned with respect to the BH spin, resulting in a tilted accretion disk. Rotation of the BH drags the surrounding space-time, manifesting as Lense-Thirring torques that lead to disk precession and warping. We study these processes by simulating a thin ($H/r=0.02$), highly tilted ($\mathcal{T}=65^\circ$) accretion disk around a rapidly rotating ($a=0.9375$) BH at extremely high resolutions, which we performed using the general-relativistic magnetohydrodynamic (GRMHD) code \verb|H-AMR|. The disk becomes significantly warped and continuously tears into two individually precessing sub-disks. We find that mass accretion rates far exceed the standard $\alpha$-viscosity expectations. We identify two novel dissipation mechanisms specific to warped disks that are the main drivers of accretion, distinct from the local turbulent stresses that are usually thought to drive accretion. In particular, we identify extreme scale height oscillations that occur twice an orbit throughout our disk. When the scale height compresses, `nozzle' shocks form, dissipating orbital energy and driving accretion. Separate from this phenomenon, there is also extreme dissipation at the location of the tear. This leads to the formation of low-angular momentum `streamers' that rain down onto the inner sub-disk, shocking it. The addition of low angular momentum gas to the inner sub-disk causes it to rapidly accrete, even when it is transiently aligned with the BH spin and thus unwarped. These mechanisms, if general, significantly modify the standard accretion paradigm. Additionally, they may drive structural changes on much shorter timescales than expected in $\alpha$-disks, potentially explaining some of the extreme variability observed in active galactic nuclei. 
\end{abstract}


\section{Introduction}
\label{sec:intro}

The traditional description of an accretion disk is the axisymmetric, geometrically thin Shakura-Sunyaev model (\citealp{ss73}; \citealp{pringle_1981} for a review; \citealp{nt73} for the relativistic treatment). While the Shakura-Sunyaev model has found widespread utility in the field, it is by no means a complete description of accretion. One of its major assumptions is that the disk angular momentum is aligned with the BH spin. Yet, in most natural circumstances, the infalling gas that forms the disk has no prior knowledge of the BH spin orientation. Thus, many accretion disks may be at least initially misaligned. This can drastically alter the dynamics of the disk because the general-relativistic `frame-dragging' of the Kerr BH will apply Lense-Thirring (LT) torques to the disk \citep{lt18,MTW}. These torques induce differential precession about the BH spin vector and can lead to large-scale warps in the disk. Early analytic work found that these LT torques can align inner regions of tilted accretion disks with the BH spin \citep{bp75}, which has sometimes been invoked to neglect the effects of disk tilt. While recent numerical simulations have confirmed the existence of BP alignment in misaligned disks with small tilts \citep{nelson_papaloizou_2000, lodato_pringle_2007, perego_2009, nealon_2015, liska_2019}, at larger tilts the story changes dramatically. In both smoothed-particle hydrodynamic (SPH) \citep{nixon_king_2012,nixon_king_2013,raj_2021,drewes_2021} and general-relativistic magnetohydrodynamic (GRMHD) simulations of highly tilted disks \citep{HAMR,liska_2021,gibwa_2022}, the LT torques are strong enough to sometimes rupture the accretion disk, splitting it either into individually precessing annuli or into discrete sub-disks. It is these highly tilted, warped and torn disks that we focus on in this paper. 

There is a wealth of analytic work devoted to understanding the dynamics of warped accretion disks. This includes early work in the linearized domain of small warps (\citealp{papaloizou_pringle_1983,kumar_pringle_1985}, see also \citealp{pringle_1992}); the fully nonlinear one-dimensional theory that generalized the study of warped accretion disks to arbitrarily sized warps \citep{ogilvie_1999,ogilvie_2000,ogilvie_latter_2013}; and more recently, the more sophisticated affine model that is general to both warps and eccentricities and treats the disk as a composition of mutable fluid columns \citep{ogilvie_affine}.

While analytic work provides a firm foundation for the understanding of warped disks, there remains only partial agreement between theory and the results of numerical simulations. In particular, disk tearing is a highly nonlinear process that results in discontinuities in the accretion flow, which are difficult to study analytically. These systems also feature anomalously high mass accretion rates that are also difficult to reconcile with theory. This was first reported in works based on SPH simulations \citep{nixon_king_2012}, which attributed the rapid accretion to the cancellation of misaligned angular momentum in torn regions. Rapid accretion is also found in GRMHD simulations of thin, tilted disks, which have reported effective viscosities well in excess of those expected in aligned thin disks \citep{liska_2021}. 

In this work, we reveal multiple novel mechanisms that enable rapid accretion in GRMHD simulations of thin, tilted disks. 
In \secref{sec:approach}, we present the details of our simulation. In \secref{sec:results}, we examine the structure of the warped accretion flow. In \secref{sec:accretion}, we show that mass accretion occurs anomalously fast, and investigate the dissipation mechanisms that drive this rapid accretion. In \secref{sec:discussion}, we contextualize the theoretical impact of our results, discuss their observational implications, and then summarize our findings.





\section{Simulation Details}
\label{sec:approach}

In this paper, we study a simulation of a thin, tilted accretion disk performed with the GPU-accelerated, 3D GRMHD code \verb|H-AMR| \citep{HAMR}. We work in spherical polar coordinates ($r$, $\theta$ and $\varphi$) and use a rapidly rotating ($a=0.9375$) black hole. We  initialize the disk with an aspect ratio $H/r=0.02$ and set the inner and outer radii to $r=6.5\,\rg$ and $r=76\,\rg$, respectively. We initialize the velocity as circular everywhere, set the radial surface density profile to $\Sigma\propto r^{-1}$, and set the vertical density profile to a Gaussian profile with a full-width at half maximum (FWHM) equal to the local scale height of the disk. We then tilt the disk with respect to the equatorial plane by $65^\circ$. We insert into the disk a purely toroidal magnetic field described by a covariant vector potential, $A_\theta \propto(\rho-0.0005)r^2$, where $\rho$ is the fluid frame gas density which is normalized such that ${\rm max\,}\rho=1$. We normalize the magnetic field so that the average ratio of gas to magnetic pressure is $\beta\approx7$, such that the disk remains dominated by gas pressure. We maintain the disk thickness by using a cooling function that removes excess internal energy from the disk \citep[e.g., ][]{noble_2009,liska_2019}. 

We perform the simulation on a spherical grid that is uniform in ${\rm log}r$ at extremely high resolutions, which are needed to resolve the turbulent motions within the disk. To achieve high resolution, we use several numerical speed-ups, including acceleration on GPUs, 3 levels of adaptive mesh refinement (AMR), and 5 levels of local adaptive time-stepping (LAT). The maximum effective resolution at $\sim10\,\rg$ in the disk, where $\rg\equiv GM/c^2$ is the gravitational radius, is $N_r\times N_\theta\times N_\phi = 13440\times4608\times8192$ cells. This resolution remains uniform within 4 disk scale heights to ensure that the disk structure is independent of the AMR criterion used. The 3 AMR levels used to achieve this resolution are added at $2$, $4$ and $8\,\rg$, such that at the event horizon the resolution is reduced to $1728\times576\times1024$ in order to prevent the minimum time-step from becoming too small, thereby speeding up the computation. To prevent the Courant condition \citep{courant_1953} in the $\varphi$ direction from limiting the time-step, we reduce the azimuthal resolution progressively from $N_\phi=1024$ cells near the equator to $16$ cells within $30^\circ$ of either pole. Both inner and outer radial boundary conditions allow matter and magnetic fields to freely leave the domain. We set the inner radial boundary to be five cells within the event horizon and the outer radial boundary to be sufficiently large such that both are causally disconnected from (and thus do not affect) the flow. The polar boundary condition is transmissive and the azimuthal boundary condition is periodic \citep{liska_2018}. We refer the reader to \cite{HAMR} for a full description of \verb|H-AMR| and \cite{gibwa_2022} for an analysis of this simulation in the context of quasi-periodic oscillations in X-ray binaries.

In a companion paper, Liska et al 2022 in prep (which we will refer to as L22), we perform the same simulation that we have analyzed here, except we separately evolve the electron and ion entropies and use an M1 closure scheme for radiation rather than a predefined cooling function. The inclusion of radiation breaks the self-similarity of the flow, so in L22 we set the black hole mass to $10\,M_\odot$ and set the Eddington ratio to $\sim0.35$. The results of the present work generally carry over in L22, and we refer the reader there for detailed comparisons of radiative versus cooled simulations of our disk. 

\begin{figure}
    \centering
    \includegraphics[width=1.1\columnwidth]{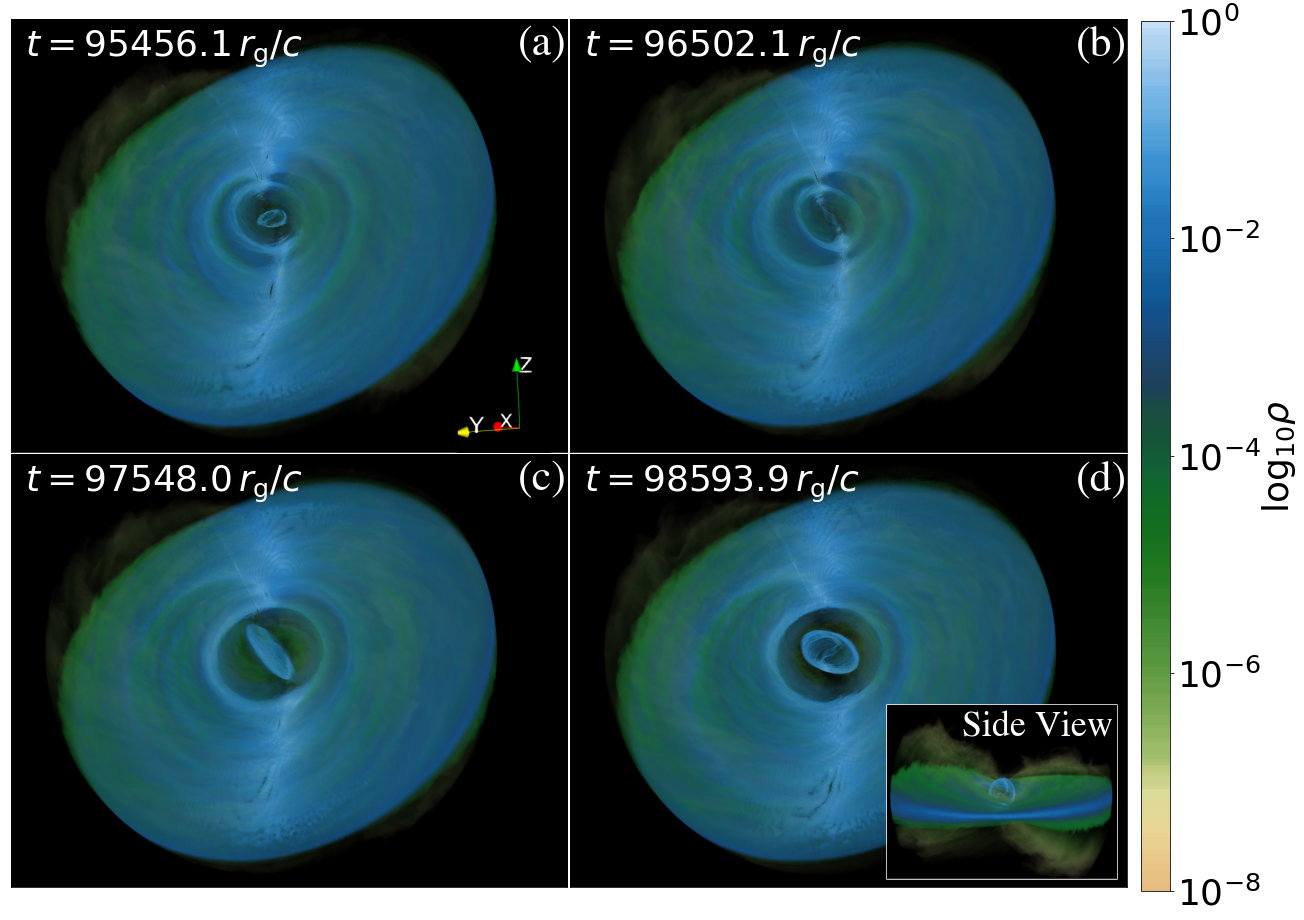}
    \caption{The Lense-Thirring torques induced by the rotation of the central black hole causes the accretion disk to warp and, sometimes, tear into discrete sub-disks. In each panel, we plot a three-dimensional rendering of the fluid frame gas density, separated by $\delta t\approx1000\,\rg/c$. Azimuthal oscillations in the scale height are apparent in the outer sub-disk, where orbiting fluid parcels experience compressions and expansions twice an orbit (evidenced by the light-blue `spokes' in the outer sub-disk). These oscillations are also apparent in the side on view of the disk shown in the inset of panel (d).}
    \label{fig:3D}
\end{figure}

\section{Accretion Geometry}
\label{sec:results}

\subsection{Main Features}
\label{sec:geometry:main}

Figure \ref{fig:3D} shows a sequence of three-dimensional renderings of the fluid frame gas density ($\rho$) in our simulation, separated in time by $\delta t\approx1000\,\rg/c$. This Figure depicts some of the main features of our simulation. The disk extends out to roughly $\approx 100\,\rg$. As the central black hole rotates, it drags space-time with it, causing the surrounding matter to rotate as well. This so-called `frame-dragging' effect induces Lense-Thirring torques \citep{lt18} onto orbiting fluid parcels. These torques cause particle orbits to precess with an angular frequency following the radial dependence $\Omega_{\rm LT}\propto r^{-3}$. Were there no (magneto-) hydrodynamic stresses acting on the disk, the disk would shred completely into independently precessing annuli. In reality, MHD stresses work to redistribute disk angular momentum. When MHD forces win, the disk maintains a warped structure, which we characterize as a series of concentric annuli with smoothly varying tilt and precession angles. However, when the Lense-Thirring torques induce a strong enough warp, the disk becomes unstable \citep{dogan_2018,dogan_2020}, leading to a runaway increase in the amplitude of the warp that manifests as a `tear'. This is seen most prominently in panels (b) and (c) of
Fig. \ref{fig:3D}, where the outer and inner regions of the disk break apart. These two `sub-disks' precess almost independently of one another. The long-lived outer sub-disk undergoes near rigid-body rotation, while the short-lived inner sub-disk often precesses differentially and is typically quickly consumed by the BH. Afterwards, the outer sub-disk refills the inner region, and the tearing process repeats.   

\begin{figure}
    \centering
    \includegraphics[width=\textwidth]{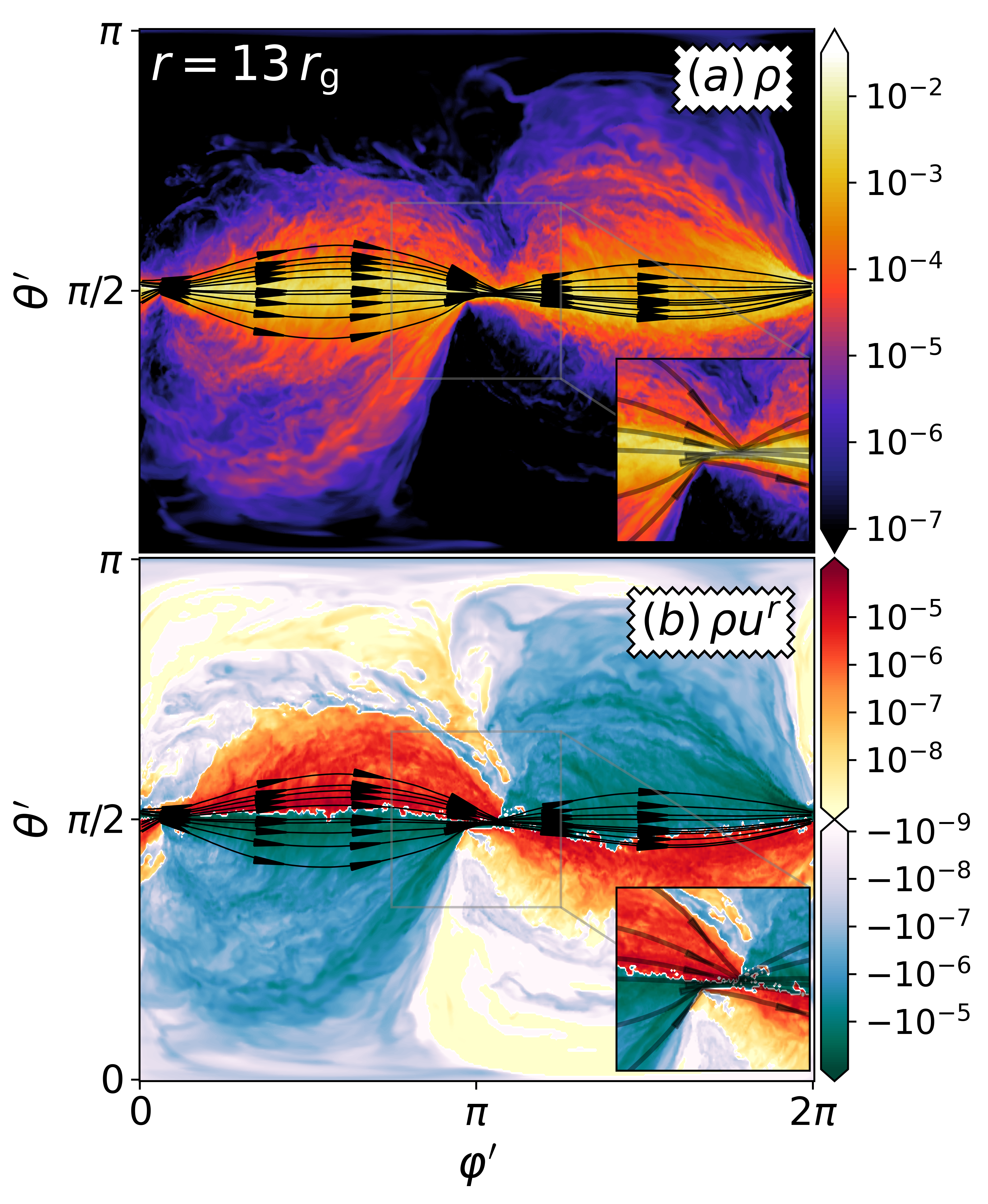}
    \caption{An annulus of the warped disk experiences vertical and radial oscillations twice an orbit. \textbf{Panel (a).} We depict the fluid frame gas density, $\rho$, at radius $r=13\,\rg$ and time $t = 90471.8\,\rg/c$. We have also drawn velocity streamlines of orbiting fluid parcels in black. The plot is depicted in tilted coordinates, $\varphi^\prime$ and $\theta^\prime$, where $\varphi^\prime=0$ indicates the local precession angle of the disk and $\theta^\prime=\pi/2$ indicates the local midplane of the disk. \textbf{Panel (b).} Same as the top panel, except we plot the radial mass flux $\rho u^r$. The radial mass flux also exhibits oscillations twice an orbit, except they are anti-symmetric about the local midplane of the disk.}
    \label{fig:flowNozzle}
\end{figure}

\begin{figure*}
    \centering
    \includegraphics[width=\columnwidth]{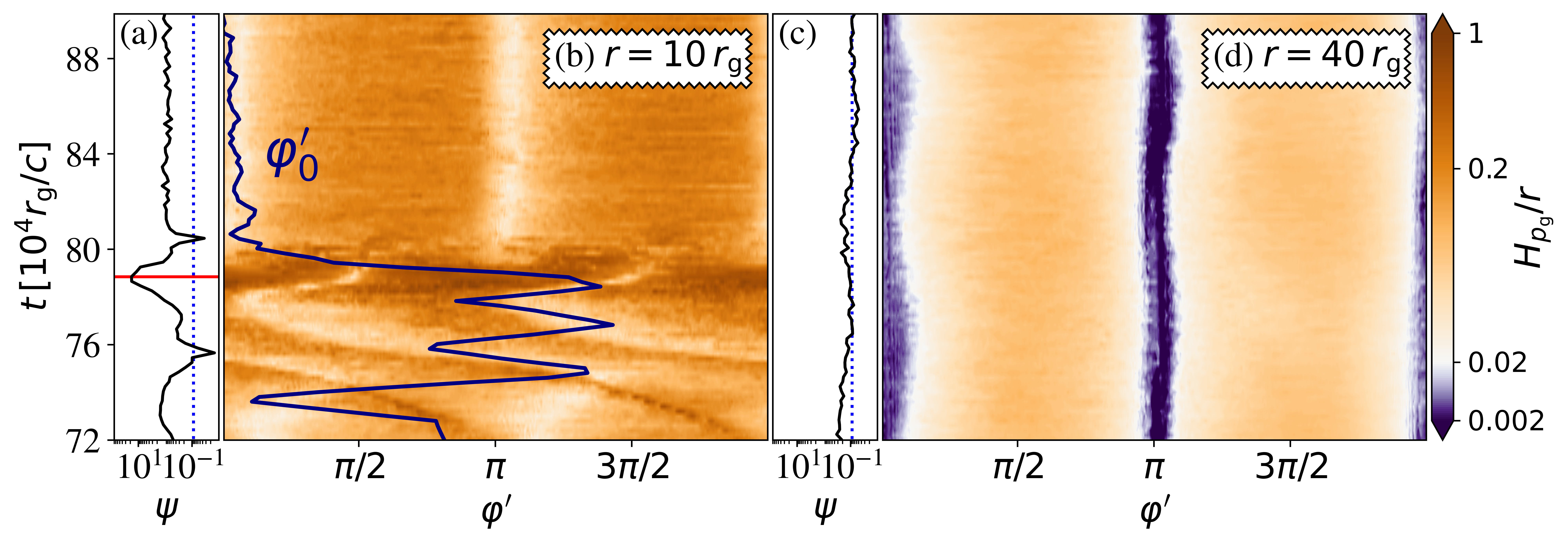}
    \caption{The phase offset of the $m=2$ scale height oscillations is equal to the precession angle (set to $\varphi^\prime=0$) in the outer disk but not in the inner disk. We depict this with space-time diagrams of $H_{p_{\rm g}}/r$ at fixed $r=10\,\rg$ (panel b) and $r=40\,\rg$ (panel d). In the time period visualized, the tearing radius is initially $10\,\rg<r_{\rm tear}<40\,\rg$, but drifts inwards, crossing $r=10\,\rg$ at $t\approx79\times10^4\,\rg/c$. This is reflected in the warp amplitude ($\psi$) plots accompanying each space-time diagram. In panel (a), we have drawn a red line when the disk tears. We have also drawn a blue dotted $\psi_{\rm c}=0.089$ line as an estimate for when extreme scale height oscillations begin occurring, calculated using Equation 154 of \cite{fairbairn_2021b}. When $t<79\times10^4\,\rg/c$, panel (b) shows the inner disk, which exhibits radially-dependent phase offsets (depicted in dark blue) due to strong differential precession. At all times at $r=40\,\rg$, the depicted annulus is part of the outer disk, and the phase of the scale height oscillations is locked with the precession angle. At $r=10\,\rg$, the scale height is also generally larger than the target scale height ($=0.02$, depicted in white) due to the enhanced dissipation and because vertical oscillations cause a departure from hydrostatic equilibrium.}
    \label{fig:HR_Spacetime}
\end{figure*}

In total, we identified  $10$ tearing events over the course of our simulation \citep[e.g., Table 1 of ][]{gibwa_2022}. A typical tearing cycle can last anywhere between $\sim10^2$ and $\sim10^4$ $\rg/c$, and while the exact location of the tear varies, it usually occurs at $r\lesssim10-20\,\rg$. The snapshots in Fig. \ref{fig:3D} depict the transition between two tearing cycles. In Fig. \ref{fig:3D} (a), the inner sub-disk from the previous tearing cycle is about to be completely consumed. In Fig. \ref{fig:3D} (b), the inner sub-disk is replenished, and is about to tear again. In Fig. \ref{fig:3D} (c), the inner sub-disk tears off once again, and continues to precess in Fig. \ref{fig:3D} (d). 

Fig. \ref{fig:3D} shows that the azimuthal distribution of gas density exhibits a periodic structure: $\rho$ increases and decreases twice an orbit, as evidenced by the twin light-blue high-density `spokes' in the outer sub-disk. Whereas the volumetric density increases locally, the surface density does not: this is because the volumetric density increase is due to vertical 
(transverse to the disk) compression. This is also shown in the inset panel of Fig. 1 (d), where we show a side-on view of the disk. This might come as a surprise, as accretion disks are typically treated as axisymmetric, without any variation in the azimuthal direction. This non-axisymmetry is fundamentally due to the warp, which we investigate in the following subsection. 

\subsection{Warp and Non-axisymmetric Flow Structures}
\label{sec:geometry:oscillations}

To better understand the non-axisymmetric structures in our accretion disk, we begin by examining the local properties of a single annulus. For this, we use tilted coordinates $r$, $\theta^\prime$ and $\varphi^\prime$, which essentially `flatten' the warp of the disk and make analysis more convenient. We describe the transformation to tilted coordinates in Appendix \ref{app:misaligned_analysis}, with an accompanying visualization in Fig. \ref{fig:app:tilt_algo}. The main features to note are,
\begin{enumerate}
    \item Because the transformation depends on the radial tilt and precession profiles, it is different at different radii. 
    \item The tilted vertical unit vector, $\hat{z}^\prime$, is co-aligned with the angular momentum of the disk at every radius.
    \item The tilted polar coordinate, $\theta^\prime$, is set such that the local mid-plane of the disk is at $\theta^\prime=\pi/2$.
    \item The tilted azimuthal coordinate, $\varphi^\prime$, is set such that the local precession angle is at $\varphi^\prime=0$.
\end{enumerate}

In Figure \ref{fig:flowNozzle}, we depict the fluid frame gas density ($\rho$, panel a) and mass flux ($\rho u^r$, panel b) as a function of $\theta^\prime$ and $\varphi^\prime$. This plot is shown at time $t = 90471.8\,\rg/c$ and radius $r=13\,\rg$. Since gas motion is predominantly azimuthal, the flow approximately follows the $\theta^\prime-\varphi^\prime$ plane. In Fig. \ref{fig:flowNozzle} (a), we can immediately see that the disk scale height varies drastically as a function of $\varphi'$. Specifically, it undergoes periodic oscillation, compressing and expanding twice an orbit. The velocity streamlines (depicted in black) also follow this periodic motion, converging and diverging in phase with the oscillations of the disk scale height. The radial mass flux, depicted in Fig. \ref{fig:flowNozzle} (b), also oscillates in phase with the scale height oscillations, reversing direction at the compression points. Additionally, these radial motions are approximately anti-symmetric about the mid-plane. Fluid parcels below (above) the midplane move inward (outward) during the $\sim0<\varphi^\prime<\pi$ expansion and outward (inward) during the $\sim\pi<\varphi^\prime<2\pi$ expansion. This radial `sloshing' of the disk is much larger in magnitude than the average, inwards radial mass flux associated with accretion. 

These oscillations can be qualitatively understood by considering how a warp impacts the hydrodynamic force balance of the disk \citep[see also Section 4.1 of ][ for a useful description]{lodato_pringle_2007}. Consider two adjacent annuli that, when unwarped, are in force equilibrium both radially and vertically. Then, impose a small tilt on one annulus with respect to the other, such that they are slightly misaligned. As fluid parcels in the two annuli orbit, at two points along the orbit they are maximally separated and at two points along the orbit they are minimally separated. In the frame of the fluid parcel, this manifests as a radial and vertical pressure gradient that oscillates twice an orbit. These pressure gradients induce corresponding oscillations of the particle orbits at the radial and vertical epicyclic frequencies, which far from the BH are approximately Keplerian. When the internal stresses of the disk respond to these perturbations, we are left with the oscillating patterns seen in Figure 2. Specifically, the vertical oscillations manifest as $m=2$ scale height oscillations while the radial oscillations manifest as an increasing eccentricity above and below the midplane \citep[as also seen in][]{deng_2021}. The argument of periastron above and below the midplane is out of phase by $\approx180^\circ$. 

These oscillations were also seen in the thick, tilted disk simulations of \cite{fragile_blaes_2008}, and were recently computed analytically. \cite{fairbairn_2021a} developed a theory of oscillating fluid tori that can describe annuli of warped disks, and then in \cite{fairbairn_2021b} they applied this theory to nonlinear warps in inviscid, Keplerian disks. They identified a bouncing regime above a critical warp amplitude, leading to large scale height variations (Fig. 2 of \cite{fairbairn_2021b}) that are remarkably similar to those seen here. We refer the reader to these works for an analytical analysis of this behavior. 


Before continuing, we must describe some of our diagnostics. This includes several averages, generally using tilted coordinates, which we define as
\begin{equation}
    \langle\cdots\rangle^x_w = \int(\cdots) W\sqrt{g_{xx}}\hat{x}\cdot d\vec{A} / \int W\sqrt{g_{xx}}\hat{x}\cdot d\vec{A},
    \label{eq:avg}
\end{equation}
where $x$ is the coordinate over which we`re performing the average and $W$ is the weight (if we use one). $g_{xx}$ is one of the components of the covariant metric tensor. We sometimes also use this notation for multiple directions, i.e., $\langle\cdots\rangle^{\theta^\prime,\varphi^\prime}_\rho$ would indicate a density-weighted average over the tilted coordinates $\theta^\prime$ and $\varphi^\prime$. 

We analyze the flow structures depicted in Figure \ref{fig:flowNozzle} by measuring the pressure scale height,
\begin{equation}
    H_{p_{\rm g}} \equiv \sqrt{\langle z^{\prime2} \rangle_{p_{\rm g}}^{\theta^\prime}},
    \label{eq:pressure_scale_height}
\end{equation}
where $p_{\rm g}$ is the fluid frame gas pressure. This expression for $H_{p_{\rm g}}$ returns the exact scale height, $H$, for an isothermal thin accretion disk in vertical hydrostatic equilibrium (i.e. for the vertical density and pressure profile $\propto {\rm exp}(-z^2/2H^2)$).

We explore the time-dependent structure of the scale height oscillations in Figure \ref{fig:HR_Spacetime}, where we show space-time ($\varphi^\prime-t$) diagrams of $H_{p_{\rm g}}/r$ at fixed radii $r=10\,\rg$ (panel b) and $40\,\rg$ (panel d). These diagrams are shown from times $t\approx72-90\times10^4\,\rg/c$, during which the inner disk is shrinking and
the tearing radius is decreasing. We have depicted our target scale height of $H/r=0.02$ with white colors, such that compressed regions are purple and expanded regions are orange. We note, however, that since the `target thickness' of the disk assumes vertical hydrostatic equilibrium, our cooling prescription is a function only of the temperature profile and does not consider vertical oscillations. Additionally, it occurs on a Keplerian timescale, effectively averaging cooling over the annulus. While this is an ad hoc treatment of the disk thermodynamics, we find much of the same  behavior in L22 where we self-consistently evolve radiation, reassuring us that our scale height evaluation is robust. 

At $t=79\times10^4\,\rg/c$, the tearing radius crosses $r=10\,\rg$. Thus, the annulus depicted in the left panel belongs to the inner sub-disk before this time and the outer sub-disk afterwards. Since our tilted coordinate system sets $\varphi^\prime=0$ to the precession angle at every radius, it is physically meaningful to analyze the phase offsets of our scale height oscillations. We derive the phase offsets by fitting $H_{p_{\rm g}}(r,\varphi^\prime)$ (Eq. \ref{eq:pressure_scale_height}) to a sinusoidal dependence that we define at a given radius as,
\begin{equation}
    \Tilde{H}_{p_{\rm g}} = \Tilde{H}_{p _{\rm g}}^{(\rm amp)}{\rm sin}(m \varphi^\prime + \varphi^\prime_0) + \Tilde{H}_{p_{\rm g}}^{(\rm mean)}
    \label{eq:fit_pressure_scale_height}
\end{equation}
While we only use our fit for $\varphi_0^\prime$ in Fig. \ref{fig:HR_Spacetime}, we will return to this expression in \secref{sec:accretion:nozzleshocks}. Fig. \ref{fig:HR_Spacetime} (b) highlights the phase offset of the depicted annulus in blue. In the inner sub-disk, the offset between the precession angle and the scale height oscillations varies in time, while in the outer sub-disk the phase offset is constant. This is also clear in the annulus depicted in Fig \ref{fig:HR_Spacetime} (d), which at all times belongs to the outer sub-disk and has a roughly constant $\varphi^\prime$ dependence. It's interesting that these offsets are constant because it implies that they evolve with the warp, i.e., $\varphi^\prime_0 \propto e^{i\omega_{\mathcal{P}}}$, where $\omega_{\mathcal{P}}(r)$ is the precession rate, which is nearly uniform in the outer sub-disk. Yet, $\varphi^\prime_0$ is a function of the warp, and points approximately at the `local' line of nodes between misaligned adjacent annuli (e.g., it is approximately pointed to by the unit vector $\hat{l}\times\partial\hat{l}/\partial r$, where $\hat{l}(r)$ is the angular momentum unit vector of an annulus). This indicates that the entire outer sub-disk precesses in a rigid, yet warped, geometry \citep[as expected analytically, e.g., Sec. 6 of ][]{fairbairn_2021b}. 

\begin{figure*}
    \centering
    \includegraphics[width=\textwidth]{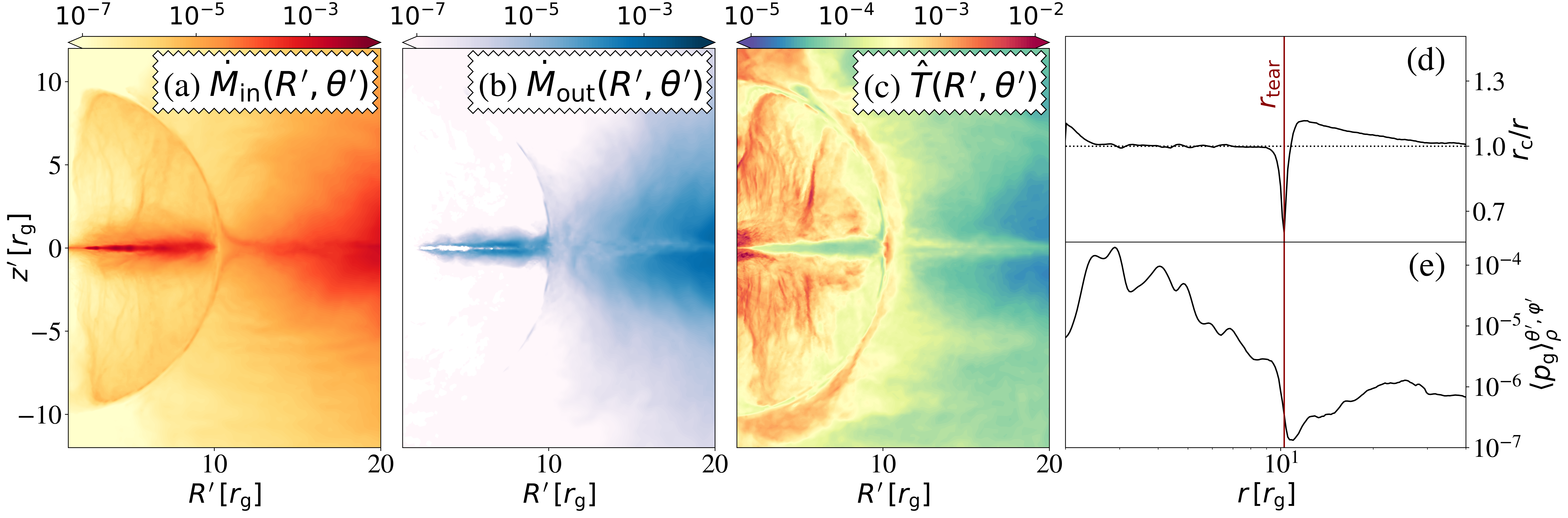}
    \caption{During a tear, the inner and outer sub-disks collide, causing streamers of low-angular gas to rain down onto the inner sub-disk. \textbf{Panel (a).} We plot contours of inward flowing mass in tilted coordinates at time $t=\,78854.3 \rg/c$. The mass flux has been integrated in $\varphi^\prime$ and is depicted in the plane of tilted cylindrical ($R^\prime$) and vertical ($z^\prime$) coordinates. At the tearing radius ($r_{\rm tear}\approx10\,\rg$) the mass flux is distributed roughly uniformly over a shell. Within the tear, gas plunges radially onto the inner disk, shocking it and increasing the disk mass. \textbf{Panel (b).} Same as panel (a), except we depict the outward flowing mass flux. \textbf{Panel (c).} Same as panel (a), except we plot the temperature of the gas. The streamer-populated region within the tear is roughly two orders of magnitude hotter than the inner sub-disk. \textbf{Panel (d).} Here, we plot the radial profile of the dimensionless circularization radius of the gas $\equiv r_{\rm c}/r$. Values above (below) unity indicate that the orbital velocity is above (below) its value for a circular orbit. At $r_{\rm tear}$, $r_{\rm c}/r$ is small, causing streamlines to plunge. Just above the tear, $r_{\rm c}/r$ is above unity. \textbf{Panel (e).} Here, we plot the radial profile of the gas pressure in the disk. Since the disk is depleted of gas at the tear, the gas pressure must have a positive radial gradient at $r\gtrsim r_{\rm tear}$. This causes an inwards force that is compensated by the super-Keplerian motion in this region. }

    \label{fig:PlungingInstantaneous}
\end{figure*}   

To support our statements about the tear, in Fig. \ref{fig:HR_Spacetime} (a) and (c) we depict the warp amplitude, $\psi\equiv |\frac{\partial \hat{l}}{\partial {\rm ln}r}|$, for each space-time diagram. In the left $\psi$ plot, we have drawn a red line at $t\approx79\times10^4\,\rg/c$ to indicate when the disk tears. We have also drawn blue dotted lines at $\psi_{\rm c}=0.089$, which is an estimate of the `critical' warp amplitude above which the extreme scale height oscillations activate. We obtained this value using Equation 154 of \cite{fairbairn_2021b} for our target scale height of $H/r=0.02$. Fig. \ref{fig:HR_Spacetime} (c) show that the $\psi\gtrsim\psi_{\rm c}$ criterion is marginally satisfied in the outer sub-disk at $r=40\,\rg$. This is a relatively mild warp, suggesting that it may be easy to activate large scale height oscillations even in disks with initial tilt angles that are much smaller than the $65^\circ$ angle considered here. 

\subsection{Tearing Region}
\label{sec:geometry:tearing}

When the disk tears, the inner and outer sub-disks begin to precess independently. As the two sub-disks evolve, they expand until they interact. This interaction is strongest at the line of nodes, where gas parcels orbiting in misaligned planes collide. These gas parcels shock, leading to significant dissipation (discussed later in \secref{sec:accretion:plunging}). This dissipation partially cancels the angular momentum of colliding gas parcels, leading to the formation of low-angular momentum streams of gas that then fall radially inwards. We refer to these as `streamers' and they are an important feature of the inner accretion flow. To better examine the flow in this region, we will define the following diagnostics. 


First, we split the mass flow into inward and outward components, 
\begin{equation}
   \dot{M}_{\rm in}(r,\theta^\prime) \equiv \int \sqrt{-g}\rho u^r \Theta(u^r)d\varphi^\prime
\end{equation}
\begin{equation}
    \dot{M}_{\rm out}(r,\theta^\prime) \equiv \int \sqrt{-g}\rho u^r \Theta(-u^r)d\varphi^\prime
\end{equation}
where $\Theta(x)$ is the Heaviside step function. We also define the average temperature of the flow as,
\begin{equation}
    \hat{T}(r,\theta^\prime) = \langle p_{\rm g} \rangle^{\varphi^\prime}/\langle \rho \rangle^{\varphi^\prime},
    \label{eq:temperature}
\end{equation}
where we have used a $`\,\hat{}\,'$ as a reminder that the units of $\hat{T}$ are non-physical. This expression also assumes that the flow is gas-pressure dominated. We expect this is true for the coronal regions, which are generally optically thin, but not for the disk. In Figure \ref{fig:PlungingInstantaneous}, we depict $\dot{M}_{\rm in}$, $\dot{M}_{\rm out}$, and $\hat{T}$ in the $R^\prime-z^\prime$ plane (where $R^\prime$ is the cylindrical radius in tilted coordinates) at time $t=78854.3\,\rg/c$. We have chosen this time because this is when the inner sub-disk is transiently aligned with the BH spin. The inner sub-disk then has zero warp and thus no scale height oscillations, allowing us to isolate the effects of the streamers from the tear on the inner sub-disk. 

In Fig. \ref{fig:PlungingInstantaneous} (a), we can see that $\dot{M}_{\rm in}$ is uniformly distributed about the tearing radius $r_{\rm tear}\approx10\,\rg$. This is because most of the angular momentum at this radius is dissipated, causing the gas distribution to spread more evenly over a spherical shell, instead of being confined to an annulus. This low angular momentum gas then forms streamers that rain down onto the inner sub-disk, seen as radially-extended filamentary structures that sandwich the inner sub-disk. Fig. \ref{fig:PlungingInstantaneous} (b) shows that $\dot{M}_{\rm out}$ follows the structure of $\dot{M}_{\rm in}$ at the tearing radius. This is a consequence of angular momentum conservation; since low angular momentum streams fall inwards, there must also be an outward transport of angular momentum. Since transport by magnetic fields is subdominant (discussed later in \secref{sec:accretion:question}), angular momentum must be carried outwards by mass. Inside the tear, where there are free-falling streamers, there is essentially zero outwardly flowing gas, as expected. In Fig. \ref{fig:PlungingInstantaneous} (c), we see that the streamer-populated regions within the tear are roughly two orders of magnitude hotter than the outer disk. This is essentially because they evolve on a dynamical timescale, making it difficult for them to cool. They also collide with the inner disk, causing them to shock, heating this region further. The hot, streamer-filled atmosphere is reminiscent of the usual corona that surrounds many accretion disks, suggesting that streamers may lead to hard X-ray emission.

Moving on to the remaining panels of Fig. \ref{fig:PlungingInstantaneous}, we first define the specific angular momentum of a particle on a circular orbit at $r=r_{\rm c}$ in a Kerr metric as \citep{shapiro_teukolsky_1983},
\begin{equation}
    \ell_{\rm c} \approx \frac{r_{\rm c}^2-2a\sqrt{r_{\rm c}}+a^2 }{\sqrt{r_{\rm c}}(r_{\rm c}^2-3r_{\rm c}+2a\sqrt{r_{\rm c}})^{1/2}}
    \label{eq:uphi_lcirc}
\end{equation}
At every radius, we calculate the BH spin aligned specific angular momentum of the gas, $\ell_c = -\langle u_{\varphi}/u_t\rangle_{\rho}^{\theta^\prime,\varphi^\prime}$, and numerically invert Eq. \ref{eq:uphi_lcirc} to find the corresponding circularization radius of the gas. We express this dimensionlessly as $r_{\rm c}/r$ and plot it as a function of radius in Fig. \ref{fig:PlungingInstantaneous} (d). At most radii, $r_{\rm c}/r\approx 1$, and the orbital motion is approximately circular. However, at $r_{\rm tear}\sim10\,\rg$ (shown in red), $r_{\rm c}$ drops substantially. This is caused by the cancellation of misaligned angular momenta. The gas at this radius can fall inwards until it reaches its local value of $r_{\rm c}$ and form streamers in the process. 

Figure \ref{fig:PlungingInstantaneous} also shows that just outside $r_{\rm tear}$, we have $r_{\rm circ}/r>1$; this means the gas there is super-Keplerian. We can understand why by turning to Fig. \ref{fig:PlungingInstantaneous} (e), where we depict the gas pressure in the disk, 
$\langle\pg\rangle_\rho^{\theta^\prime,\varphi^\prime}$, as a function of radius. The depletion of gas at $r_{\rm tear}$ causes a dip in pressure, which results in a positive pressure gradient at radii $\gtrsim r_{\rm tear}$. This results in a pressure force that is pointed inwards, which is what compensates the super-Keplerian centrifugal force of the gas.




\section{Accretion Mechanisms}
\label{sec:accretion}

\subsection{Why do highly tilted accretion disks accrete so rapidly?}
\label{sec:accretion:question}

In the previous section, we focused on the structure of the warped accretion flow in our simulations. Now, we will study how this structure determines the accretion mechanisms in our disk. In aligned disks, angular momentum transport is usually parameterized by the $\alpha$ parameter, which sets the strength of an effective viscosity and is, in reality, thought to represent magnetized turbulence driven by the magnetorotational instability \citep[MRI, ][]{balbus_hawley_1981}. It is theoretically expected that $\alpha<1$. This is because the turbulence should be subsonic and confined by the scale height of the disk; i.e., viscosity is $v \approx l_{\rm eddy}V_{\rm eddy} \approx \alpha Hc_{\rm s}$ where $l_{\rm eddy}<H$ and $V_{\rm eddy}<c_{\rm s}$ are the characteristic length and velocity scales of the eddies \citep{pringle_1981}. 

\begin{figure}
   \centering
    \includegraphics[width=\columnwidth]{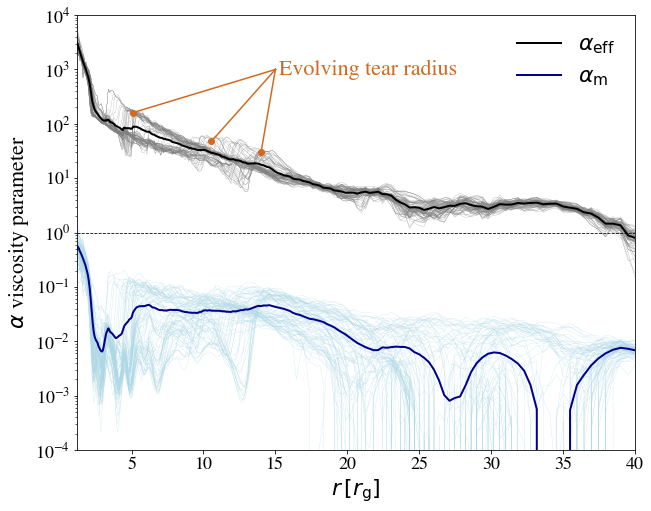}
    \caption{The effective ${\alpha}$ parameter far exceeds both unity and the magnetic $\alpha$ parameter, suggesting that accretion is neither driven by local turbulence (where $\alpha\lesssim1$) nor magnetic fields. We show this by plotting radial profiles of the effective $\alpha$ parameter, $\alphaeff$ (Eq. \ref{eq:alpha_eff}), and the $\alpha$ parameter associated with the Maxwell's stress, $\alpham$ (Eq. \ref{eq:alpha_maxwell}). These profiles are averaged from times $\sim72-90\times10^4\,\rg/c$. For both quantities, we also plot their instantaneous profiles at each time, which are depicted in lighter colors. Our time-averages are not true steady-state measures of $\alpha$ because disk tearing makes the accretion flow is inherently transient. The imprint of disk tearing can be seen in the instantaneous $\alpha_{\rm eff}$ curves, where we have labeled the radius of the tear, which moves inwards as the inner disk is accreted. }
    \label{fig:flowAlpha}
\end{figure}

Highly tilted and warped accretion disks can, however, accrete at much higher rates (see the reported effective $\alpha$ parameters in \citealp{liska_2021}, or the dynamically driven accretion reported in \citealp{nixon_king_2012}). Figure \ref{fig:flowAlpha} shows the radial profiles of the effective $\alpha$ parameter, $\alphaeff$, and the $\alpha$ parameter derived from the Maxwell's stress, $\alpham$. Both quantities are averaged over the duration of a tearing event, $t\sim72-90\times10^4\,\rg/c$. It is instructive to first look at $\alpha_{\rm eff}$, which is the effective $\alpha$ parameter that results when assuming  mass accretion is fully driven by local viscous processes,
\begin{equation}
    \alpha_{\rm eff}= \frac{\langle \rho u^r \rangle^{\theta^\prime,\varphi^\prime,t}\langle u_{\varphi^\prime}\rangle^{\theta^\prime,\varphi^\prime,t}_\rho}{r\langle p_{\rm g}\rangle_\rho^{\theta^\prime,\varphi^\prime,t}}
    \label{eq:alpha_eff}
\end{equation}
 We can interpret $\alphaeff$ as the $\alpha$ parameter associated with local turbulent stresses if accretion is, in fact, driven by local turbulent stresses. However, as we can see from Fig. \ref{fig:flowAlpha}, $\alphaeff$ is $\approx1-100$ through much of the disk, which is much larger than the $\alpha\leq1$ theoretical limit imposed on turbulent stresses. Additionally, we can compare $\alphaeff$ to $\alpham$, which is defined as, 
\begin{equation}
    \alpha_{\rm M} = \frac{-\langle b^rb_{\varphi'}\rangle_\rho^{\theta^\prime,\varphi^\prime,t}}{r\langle p_{\rm g}\rangle_\rho^{\theta^\prime,\varphi^\prime,t}}
    \label{eq:alpha_maxwell}
\end{equation}
where $b^\mu$ is the four-vector of the fluid frame magnetic field. Here, $\alpha_{\rm M}$ represents angular momentum transport driven by magnetized turbulence, thought to be seeded by the MRI, and Fig. \ref{fig:flowAlpha} shows that it is bounded below unity as expected. However, since $\alpha_{\rm M}$ is $2-3$ orders of magnitude below $\alphaeff$, we can definitively say that magnetic stresses do \textit{not} drive accretion in our simulation.\footnote{We note the absence of an $\alpha$ parameter associated with Reynolds' stresses in Fig. \ref{fig:flowAlpha}; this was an intentional choice. The oscillating flow structures in our warped disk break the ergodicity of the disk in the $\hat{\varphi}^\prime$ direction and would merit a spectral Reynolds' decomposition of the flow into its Fourier modes, which is beyond the scope of this work.}

The main takeaways from Fig. \ref{fig:flowAlpha} are: (i) mass accretion occurs much faster in warped accretion disks than in equatorial accretion disks, (ii) the transport is likely non-local in nature, since $\alphaeff\gg1$, and (iii) accretion must occur mainly via non-magnetic stresses, since $\alpha_{\rm M}\ll\alpha_{\rm eff}$. In the following subsections, we identify the two main accretion mechanisms in our simulation.

\subsection{Nozzle Shocks}
\label{sec:accretion:nozzleshocks}
\begin{figure}
    \centering
    \includegraphics[width=\textwidth]{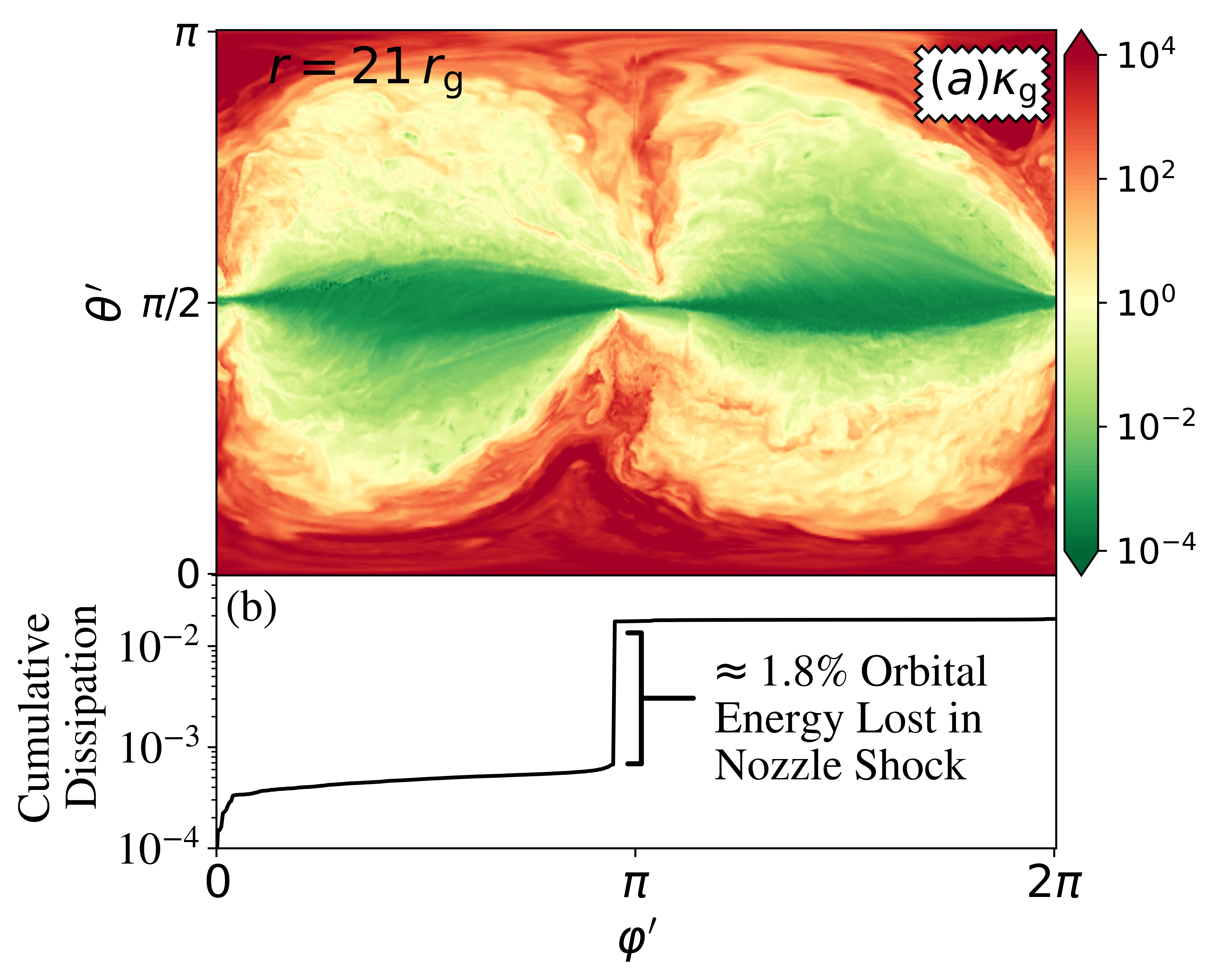}
    \caption{The vertical compressions of the warped disk lead to `nozzle shocks' twice an orbit, where significant dissipation occurs. \textbf{Panel (a).} Here, we depict the fluid frame entropy ($\kappa_{\rm g}$), at radius $r=21\,\rg$ and time $t=115923.3\,\rg/c$ in the $\theta^\prime-\varphi^\prime$ plane. We see that the entropy spikes at $\varphi^\prime\approx0$ (or $2\pi$) and $\pi$, where the disk is most compressed. \textbf{Panel (b).} We plot the cumulative fraction of dissipated energy along the annulus (Eq. \ref{eq:integrated_dissipation_rate} and following text), normalized to the orbital energy of the annulus. Across the $\varphi^\prime=\pi$ nozzle shock, $\approx1.8\%$ of the orbital energy is dissipated.}
    \label{fig:flowNozzleEntropy}
\end{figure}

\begin{figure*}
    \centering
    \includegraphics[width=\textwidth]{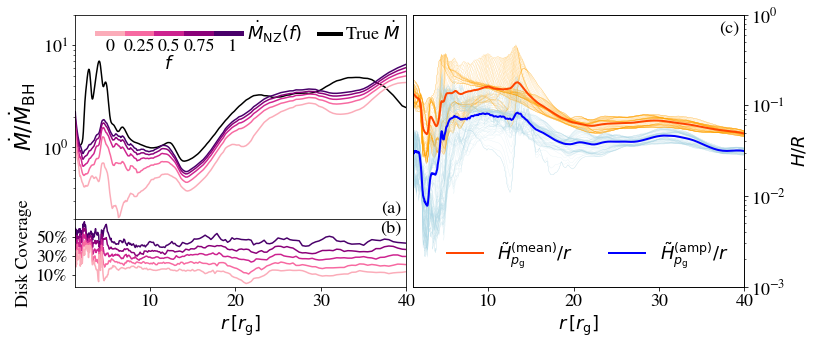}
    \caption{Dissipation in nozzle shocks can mostly account for the measured mass accretion rate everywhere except the innermost regions of the disk. \textbf{Panel (a).} Here, we compare $\dot{M}$ to the estimated mass accretion rate due to nozzle shocks, $\dot{M}_{\rm NZ}$ (Eq. \ref{eq:mdot_nz}). We do this for multiple values of $f$ (Eq. \ref{eq:Flux_Q}), where higher $f$ indicates a stricter cutoff for the degree of compression. Each quantity is averaged from times $\sim72-90\times10^4\,\rg/c$. \textbf{Panel (b).} Here, we show what percentage of a given annulus passes our compression criterion when calculating $\dot{M}_{\rm NZ}$, for each value of $f$. \textbf{Panel (c).} Here, we show our fits for the dimensionless mean ($\Tilde{H}_{p_{\rm g}}^{\rm (mean)}/r$) and amplitude ($\Tilde{H}_{p_{\rm g}}^{\rm (amp)}/r$) of the  scale height oscillations (Eq. \ref{eq:fit_pressure_scale_height}). The quantities are time-averaged over the same period as the other panels, with the corresponding instantaneous curves depicted in lighter colors.}
    \label{fig:MdotNozzle}
\end{figure*}

In Section \secref{sec:geometry:oscillations} we identified extreme scale height variations that occur twice an orbit. We will now show that these compressions lead to the formation of shocks that dissipate orbital energy. These `nozzle shocks' are conceptually similar to those that occur in tidal disruption events \citep[TDEs,][]{rees_1988,kochanek_1994}. They were also seen in the thicker tilted disk simulations of \cite{fragile_blaes_2008}, who dubbed them `standing shocks'. More recently, \cite{fairbairn_2021b} also suggested nozzle shocks may form in warped disks.

 Figure \ref{fig:flowNozzleEntropy} shows the specific entropy, $\kappa_{\rm g}=p_{\rm g}/\rho^\gamma$, in the $\theta^\prime-\varphi^\prime$ plane at radius $r=21\,\rg$ and time $t=115,923.3\,\rg/c$. In a steady, laminar flow, $\kappa_{\rm g}$ should be conserved along streamlines. However, if the gas shocks, then $\kappa_{\rm g}$ will increase, making it a useful quantity for tracking dissipation. We can see that throughout most of the depicted annulus, $\kappa_{\rm g}$ is roughly constant, suggesting that the compressions are mostly adiabatic. We say `mostly', however, because at the points of maximum compression, there is in fact dissipation occurring. To ascertain this, we start by expressing the heating rate per unit volume of a fluid parcel by its the change in entropy\footnote{Entropy is also lost via our cooling function. However, we don't need to consider it in this expression, as we are only concerned with the entropy dissipated in shocks - not the entropy that's removed by the cooling function.} \citep{ressler_2015}, 
\begin{equation}
    Q = \rho^{\gamma}(\gamma-1)^{-1}u^\mu\partial_\mu \kappa_{\rm g}, 
\label{eq:entropy_heating}
\end{equation}
In practice, we use only positive values of $Q$, since negative values result in a decrease in entropy along a streamline and are unphysical in a steady flow\footnote{While our flow is generally transient, this is a fine assumption in the $\hat{\varphi^\prime}$ direction since the orbital timescale is generally much shorter than any other timescale.}. While this is a somewhat crude fix, a more precise treatment would require a dedicated heating scheme, i.e. Section 3.2 of \cite{ressler_2015}. To analyze dissipation in our nozzles, we are concerned only with the dissipation of the $\hat{\varphi}^\prime$ velocity components, so we will also define $Q_{\varphi^\prime} \equiv \rho^\gamma(\gamma-1)^{-1}u^{\varphi^\prime}\partial_{\varphi^\prime}\kappa_{\rm g}$. We then integrate the azimuthal heating rate vertically,
\begin{equation}
    F_{\varphi^\prime} = \int_0^\pi \sqrt{g_{\theta^\prime\theta^\prime}}Q_{\varphi^\prime} d\theta^\prime
    \label{eq:integrated_dissipation_rate}
\end{equation}
Then, we perform a cumulative integral of this quantity in azimuth, $\int_0^{\varphi^\prime}F_{\varphi^\prime}d\varphi^\prime$. We plot this in the bottom panel of Fig. \ref{fig:flowNozzle}, where we have normalized our integral to the estimated orbital energy per unit area, $\Sigma \langle u_t\rangle^{\theta^\prime,\varphi^\prime}_{\rho}$, where $\Sigma\equiv \int_0^{\pi}\sqrt{g_{\theta^\prime\theta^\prime}}\rho d\theta^\prime$ is the surface density of the disk. We can then see a very sharp discontinuity almost exactly at $\varphi^\prime=\pi$ of scale $\approx0.018$, indicating that about $1.8\%$ of orbital energy is lost in the nozzle shock. Extrapolating, this would suggest that $\sim3.6\%$ of orbital energy is lost every orbit, since there are two nozzle shocks. This is a significant dissipation rate, which we will now show is enough to power the rapid accretion that we reported in \secref{sec:accretion:question}.

We would like to isolate the dissipation associated with the nozzle shocks, so we will use a criterion to select compressed regions in our integration defined in Eq. \ref{eq:integrated_dissipation_rate}. To do this, we use our sinusoidal fit to $H_{p_{\rm g}}(r,\varphi^\prime)$, defined in Eq. \ref{eq:fit_pressure_scale_height}. 
Then, we only consider the energy dissipation where $H_{p_{\rm g}} < A(f)$, where $A(f)\equiv \Tilde{H}_{p _{\rm g}}^{(\rm mean)} - f\Tilde{H}_{p _{\rm g}}^{(\rm amp)}$ and $0\leq f\leq1$. For the specific choice of $f=\frac{1}{\sqrt{2}}$, $A$ is one standard deviation below the mean of Eq. \ref{eq:fit_pressure_scale_height}. We then rewrite Eq. \ref{eq:integrated_dissipation_rate} with our $A(f)$ criterion,
\begin{equation}
    F_{\rm NZ}(r) = \int_0^\pi \sqrt{g_{\theta^\prime\theta^\prime}} Q_{\varphi^\prime} \Theta(A(f)-H_{p_{\rm g}})d\theta^\prime
    \label{eq:Flux_Q}
\end{equation}
Next, we would like to relate this to a predicted mass accretion rate that we can compare with the true mass accretion rate. To do this, we treat $F_{\rm NZ}$ as an axisymmetric, Newtonian dissipation rate associated with a shear viscosity. We can then relate $F_{\rm NZ}$ to a predicted accretion rate \citep[for details see][]{pringle_1981} by writing,
\begin{equation}
    \dot{M}_{\rm NZ} \approx \frac{4\pi r^3 F_{\rm NZ}}{3GM}
    \label{eq:mdot_nz}
\end{equation}

In Figure \ref{fig:MdotNozzle} (a), we compare $\dot{M}_{\rm NZ}$ to the simulated $\dot{M}$ for various values of $f$. Here, we have time-averaged $\dot{M}_{\rm NZ}$ over the period $\sim72-90\times10^4\,\rg/c$ (same as Fig. \ref{fig:flowAlpha}). Higher values of $f$ indicate a stricter criterion for selecting compressed regions along an annulus. We can see that for $f=0-0.75$, $\dot{M}_{\rm NZ}$ matches $\dot{M}$ within an order-unity factor everywhere except the innermost regions. At $f=1$, $\dot{M}_{\rm NZ}$ starts departing farther from $\dot{M}$ because our criterion for selecting the dissipation region becomes too strict. To aid our intuition for the `strictness' of our criterion, Fig. \ref{fig:MdotNozzle} (b) shows the percentage of the disk used in our integration at every radius for each $f$. We see that when $f=0.75$, for which $\dot{M}_{\rm NZ}$ largely accounts for $\dot{M}$ at most radii, we only select roughly $\sim20\%$ of an annulus at any given radius, which is an already rather small fraction of the disk. Note that $\dot{M}_{\rm NZ}$ underestimates $\dot{M}$ in the inner regions. This is because the streamers discussed in \secref{sec:geometry:tearing} can additionally drive accretion, which we explore further in \secref{sec:accretion:plunging}. 

Fig. \ref{fig:MdotNozzle} (c) shows our time-averaged estimates for the dimensionless scale height mean, $\Tilde{H}_{p_{\rm g}}^{(\rm mean)}$, and amplitude, $\Tilde{H}_{p_{\rm g}}^{(\rm amp)}$. For both, we also plot instantaneous curves in lighter colors to give the reader a better sense of the time-variability of the scale height. At all radii, the mean exceeds the amplitude. In general, the mean tends to be larger than our target scale height $H/r=0.02$. The reason for this is twofold; first, the enhanced dissipation rate makes cooling to target thickness more difficult, and second the cooling prescription does not account for the vertical oscillations which also inflate the scale height. Both the mean and the amplitude, however, become small at $r\lesssim5\,\rg$because the inner region aligns with the BH spin during the phase we have time-averaged, removing the warp.

\subsection{Streamers}
\label{sec:accretion:plunging}

In \secref{sec:geometry:tearing}, we showed than when our disk tears in two, the sub-disks can interact and lead to dissipation. This dissipation results in the formation of low-angular momentum streamers that rain down onto the inner sub-disk. In this subsection we will study how this process can enhance accretion in the inner region, as we saw in Fig. \ref{fig:PlungingInstantaneous}. We follow our approach in \secref{sec:geometry:tearing} where we focused on a tearing cycle in which the inner sub-disk transiently aligns with the BH spin axis. During this aligned phase, the inner disk is unwarped and thus has no nozzle shocks. This allows us to isolate the effects of streamers on the accretion process.

\begin{figure*}
    \centering
    \includegraphics[width=\textwidth]
    {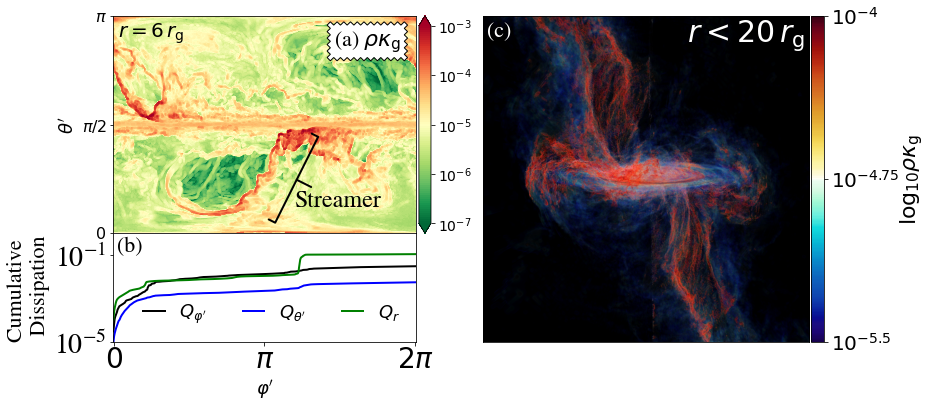}
    \caption{Low-angular momentum streamers produced at the tear crash into the inner sub-disk, shocking it and leading to significant dissipation. \textbf{Panel (a).} We depict the fluid frame entropy density ($\rho\kappa_{\rm g}$) at radius $r=6\,\rg$ and time $t=\,81,062\rg/c$. Streamers that originated at $r_{\rm tear}=7.8\,\rg$ rain down onto a transiently aligned, unwarped inner sub-disk, shocking and dissipating their kinetic energy. \textbf{Panel (c).} We depict a three-dimensional rendering of $\rho\kappa_{\rm g}$ at the same time. Here, we have excised gas at radii $>20\,\rg$ to focus on the inner regions. We can see that streamers from the outer sub-disk rain down onto the inner sub-disk from either side, and then `spill' over the top. \textbf{Panel (b).} We depict the cumulative dissipation, analogous to Fig. \ref{fig:flowNozzleEntropy} (b), except we show the dissipation in each direction ($r,\theta^\prime,\varphi^\prime$). We see that we get dissipation of a similar strength to a nozzle shock, except the dissipation is more spread out on the inner disk. The dissipation is still centered at two points along the annulus, but is no longer located at $\varphi^\prime\approx0$ and $\pi$.}
    \label{fig:plunge_entropy}
\end{figure*}

We would like to first qualitatively depict the impact of streamers on the inner sub-disk. We do this in Figure \ref{fig:plunge_entropy} (a), where we have plotted the entropy density $\rho\kappa_{\rm g}$ in the $\theta^\prime-\varphi^\prime$ plane at radius $r=6\,\rg$. This is aided by Fig \ref{fig:plunge_entropy} (c), which shows a three-dimensional rendering of $\rho\kappa_{\rm g}$, where we have excised gas at radii $r>20\,\rg$ to focus on the inner regions. Both snapshots are at time $t=81,062\,\rg/c$ when the tear is at $r_{\rm tear}=7.8\,\rg$. We find entropy density, rather than specific entropy, a useful quantity to plot as it encodes both dissipation in shocks and adiabatic compression. In both two- and three-dimensional renderings, we see that the streamers collide with the sub-disk, resulting in an increase of $\rho\kappa_{\rm g}$. After they collide, some of the material from the streamers `spills' over the inner sub-disk. 

In Fig. \ref{fig:plunge_entropy} (b), we depict the cumulative dissipation along the azimuthal direction. 
This is analogous to our calculation in Fig. \ref{fig:flowNozzleEntropy} (see also Eq. \ref{eq:integrated_dissipation_rate} and following text), except we show dissipation rates for each direction ($r$, $\theta^\prime$, and $\varphi^\prime$). We do this because the streamer trajectories are significantly altered from the mean flow of the disk. We see that as a streamer crashes into the disk, it shocks, leading to significant dissipation. If we compare this to the nozzle shock dissipation in Fig. \ref{fig:flowNozzle} (b), the dissipation is still concentrated at two points along the annulus, because at each radius streamers will make landfall both above and below the inner sub-disk. However, the dissipation is less `peaked' than in a nozzle shock, more spread in azimuth, and not concentrated near $\varphi^\prime = 0$ and $\pi$.

\begin{figure}
    \centering
    \includegraphics[width=\textwidth]{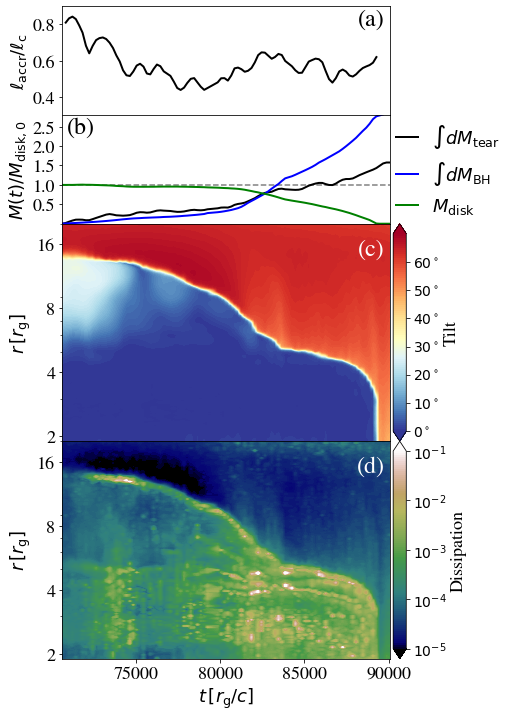}
    \caption{Dissipation at the tearing radius causes the formation of low-angular momentum streams of gas (`streamers') that plunge onto the inner sub-disk, causing it to rapidly reduce in size. \textbf{Panel (a).} We depict a time-series of the specific angular momentum of gas accreted at the tear ($\ell_{\rm accr}$, Eq. \ref{eq:l_accr}) normalized to the specific angular momentum of a circular orbit at the tearing radius ($\ell_{\rm c}$, Eq. \ref{eq:uphi_lcirc}). At all times during the depicted tearing cycle, $\ell_{\rm accr}<\ell_{\rm c}$, indicating that the gas will try to circularize at smaller and smaller radii. \textbf{Panel (b).} We show a time-series of the mass of the inner sub-disk ($M_{\rm disk}$), the mass accreted by the BH ($\int dM_{\rm BH}$, Eq. \ref{eq:mdot_bh}), and the mass accreted by the inner sub-disk from the tear ($\int dM_{\rm tear}$, Eq. \ref{eq:mdot_tear}). Each is normalized to the initial mass of the inner sub-disk during the depicted phase ($M_{\rm disk,0}$). At $t\approx83000\,\rg/c$, there is a transition, when low-angular momentum gas from the tear fully replenishes the initial mass of the inner sub-disk. \textbf{Panel (c).} Here, we plot a corresponding space-time diagram of the tilt angle. The tear is delineated by the discontinuity between red and blue regions. By $t\approx77500\,\rg/c$, the inner sub-disk fully aligns with BH spin. The inner sub-disk is consumed by the BH at the end of the depicted phase and is quickly replaced by
    outer disk material that has tilt angle $\approx60^\circ-65^\circ$. \textbf{Panel (d).} We show a space-time diagram of the shell-integrated dissipation rate ($\int Q dA_{\theta\varphi}$). There is generally dissipation at the tear and at radii $r\lesssim5-6\,\rg$, where streamers crash into the inner sub-disk. Dissipation in these regions is correlated, since dissipation in the tear is associated with an increased formation rate of streamers.}
    \label{fig:PlungingRdisk}
\end{figure}

To make this analysis more quantitative, we will invoke the conservation of mass and BH-aligned angular momentum, and track the fluxes of each entering and leaving the inner sub-disk. We start by writing down the mass of the inner sub-disk,
\begin{equation}
    M_{\rm inner} = \int_{r<r_{\rm tear}} \rho \sqrt{-g}drd\theta d\varphi^\prime
\end{equation}
We also track the mass and (BH spin aligned) angular momentum accreted by the inner sub-disk from the outer sub-disk,
\begin{equation}
    \dot{M}_{\rm tear} = \int_{r=r_{\rm tear}} \rho u^r dA_{\theta\varphi}
    \label{eq:mdot_tear}
\end{equation}
\begin{equation}
    \dot{L}_{\rm tear} = \int_{r=r_{\rm tear}} \rho u^r u_\varphi dA_{\theta\varphi}
    \label{eq:ldot_tear}
\end{equation}
We also will use the mass accreted by the BH, 
\begin{equation}
    \dot{M}_{\rm BH} = \int_{\rm event\,horizon} \rho u^r dA_{\theta\varphi}
    \label{eq:mdot_bh}
\end{equation}
Finally, we define use the specific angular momentum of the gas accreted onto the inner sub-disk as,
\begin{equation}
    \ell_{\rm accr}\equiv \dot{L}_{\rm tear}/\dot{M}_{\rm tear}
    \label{eq:l_accr}
\end{equation}
In Figure \ref{fig:PlungingRdisk} (a), we plot a time series $\ell_{\rm accr}$ normalized to the specific angular momentum of a circular orbit at $r=r_{\rm tear}$, $\equiv \ell_{\rm c}$ (Eq. \ref{eq:uphi_lcirc}). The time series is plotted over the course of the same tearing cycle depicted in Figs. \ref{fig:PlungingInstantaneous} and \ref{fig:plunge_entropy}, where the inner sub-disk transiently aligns with the BH spin. This is depicted in Fig. \ref{fig:PlungingRdisk} (c), where we plot a space-time diagram of the tilt angle, $\mathcal{T}$. Here, the discontinuity between $\mathcal{T}\approx0^\circ$ and $65^\circ$ indicates the tear. At all times during this phase, $\ell_{\rm accr}/\ell_{\rm c} \approx 0.5-0.7$. This means that all the gas the inner sub-disk accretes will try to circularize at smaller and smaller radii. If the mass of the accreted low-angular momentum gas becomes comparable to the mass of the inner sub-disk, then this will cause the inner sub-disk to shrink. 

In Fig. \ref{fig:PlungingRdisk} (b), we plot three masses normalized to the mass of the inner sub-disk at the beginning of the depicted phase, $M_{\rm disk,0}$. We show the mass of the inner sub-disk, $M_{\rm disk}$, the total mass the inner sub-disk accretes during the phase, $\int dM_{\rm tear} = \int \dot{M}_{\rm tear}dt$, and the total mass the BH accretes from the inner sub-disk, $\int dM_{\rm BH} = \int \dot{M}_{\rm BH}dt$. Before about $t\approx83,000\,\rg/c$, the BH accretion rate essentially traces the tear accretion rate, suggesting the inner sub-disk acts as a `conveyor belt' of material. Since the accreted gas from the tear has low specific-angular momentum, the inner sub-disk begins to shrink during this phase. At $t\approx 83000\,\rg/c$, each curve intersects; the inner sub-disk has been depleted of it`s initial mass by the black hole but has also been replenished by the tear. Since all of this replenished material will try to circularize at smaller radii at the time of accretion, the sub-disk must shrink. After this transition, the inner sub-disk rapidly plunges into the BH,  marking the end of the tearing cycle. 

Dissipation must be happening for the rapid accretion of the inner sub-disk to occur. We know that some of this dissipation occurs at the tear, since streamlines that pass through it lose a large fraction of their angular momentum. However, as we showed in Fig. \ref{fig:plunge_entropy}, dissipation is also occurring where these streamers merge with the inner sub-disk. To get a better sense of the positional dependence of the the dissipation rate, we show a space-time diagram of $\int Q dA_{\theta\varphi}$ in Fig. \ref{fig:PlungingRdisk} (d). We can see that along the tear, there is generally a peak in the dissipation rate, but this peak can vary in strength by about $1-2$ orders of magnitude. Correlated with this peak is a spread in the dissipation rate at radii $\lesssim5\,\rg$; these are due to the streamers. In the final stages of the tearing cycle, the streamer dissipation rate is particularly enhanced. Together, this suggests that dissipation at the tear and where the streamers collide with the inner sub-disks are comparably important. We also argue that this effect is the cause for the discrepancy between $\dot{M}$ and $\dot{M}_{\rm NZ}$ seen at radii $\lesssim5-10\,\rg$ in Fig. \ref{fig:MdotNozzle} (a), indicating that streamer- and tear- induced dissipation are essential contributors to accretion in torn disks.

\section{Discussion}
\label{sec:discussion}

\subsection{Expanding the standard model of accretion disks}
\label{sec:discussion:accretion}

The results of Section \secref{sec:accretion} have profound implications for the understanding of accretion disks. For decades, the standard picture has been that thin, magnetized disks are subject to the MRI which systematically drives angular momentum outward and mass inwards, thus enabling accretion \citep{balbus_hawley_1981}. Although we have demonstrated that magnetic stresses do contribute to accretion ($\alpha_{\rm M}$ curve in Fig. \ref{fig:flowAlpha}), they are highly subdominant to the warp-induced dissipation from nozzle shocks (\secref{sec:accretion:nozzleshocks}, disk tearing and streamers (\secref{sec:accretion:plunging}). This is not to say the MRI can be neglected in our simulations; were there no magnetized turbulence at all, we expect that the disk would immediately shred because there would be nothing to initially withstand the differential Lense-Thirring torques induced by the BH (as discussed in \secref{sec:geometry:main}). Furthermore, the accretion mechanisms studied here are strongest in the inner regions of the disk, as evidenced by the steep radial dependence in Fig. \ref{fig:flowAlpha}. Most astrophysical disks are, in reality, much larger in radial extent than the one we have simulated. Active galactic nuclei (AGN) can harbor \cite{nt73} disks up to radii $\sim10^3-10^6\,\rg$ (depending on the mass accretion rate) before becoming Toomre-unstable \citep{toomre_1964,sirko_goodman_2003,thompson_quataert_2005}. XRB disks are fundamentally limited in size by their Hill radius, which is for instance $\sim10^5-10^6\,\rg$ for estimated orbital parameters of Cygnus X-1 \citep{millerjones_2021}, although the accretion flow will likely circularize at much smaller radii. So, in misaligned disks hosted by black holes across the mass spectrum, it may be that the MRI drives accretion through most of the disk while the warp drives accretion in the inner regions. 

A difficulty of \cite{nt73} disks is that their radiation-dominated inner regions are both thermally \citep{pringle_1973} and viscously \citep{lightman_eardley_1974} unstable. This is essentially because any increase (decrease) in temperature results in an increase (decrease) in viscous heating, thus triggering a runaway process. This instability is specific to $\alpha$-disks, since it is assumed that the the viscous stress is $\sim\alpha P_{\rm tot}$ which is $\propto T^4$ for a radiation pressure dominated gas. Yet, there is little observational evidence of the thermal-viscous instability of thin disks, which appear to be stable up to significant fractions of the Eddington luminosity \citep{Done_2004}. We suggest that nozzle shock driven accretion is unlikely to be subject to this instability. We expect this because thinner disks are more compressible and thus more susceptible to dissipation in nozzle shocks. If dissipation increases, then the disk will puff up, increasing the scale height and thus decreasing the nozzle shock dissipation rate and maintaining stability. We are unable to probe the stabilizing effect of nozzle shocks in this work since we employ a predefined cooling rate. However, in L22 where we re-run our simulation using an M1 closure scheme for radiation, we find that our disk is in fact thermally stable.  

It is important to recognize that we have analyzed a single simulation in this work. Paired with the results of \cite{fragile_blaes_2008}, \cite{nixon_king_2012} and \cite{liska_2021}, it does appear that rapid accretion may be a generic feature of warped disks, but the parameter dependence must be explored before we can ascertain this. One of the most important parameters is the tilt angle. However, if accretion happens at random angles (this is the case if the BH spin and the gas supply have no prior knowledge of one another), then the average tilt angle is $60^\circ$. This is only $5^\circ$ shy of the tilt angle used here, so the accretion mechanisms we have studied may be quite general. 

Another critical parameter is the scale height of the disk. Thinner disks will be more strongly influenced by a warp, both for geometric reasons and because their efficient cooling makes them more compressible. So, if we increase the aspect ratio of our disk, the disk will be harder to tear and warps may be less efficient at driving accretion. It is possible then that the mechanisms discussed in this work are corrections to the accretion process rather than than the primary drivers of accretion. The scale height is generally set by the Eddington ratio (the ratio of the mass accretion rate of the BH to its Eddington luminosity for a given radiative efficiency), which provides us a more `astrophysical' scale for examining the aspect ratio of the disk. Disks at Eddingtion ratios of $\sim0.1-10\%$ generally cool efficiently and thus lead to thinner disks. At higher Eddington ratios, radiation pressure begins dominating, puffing the disk up. At lower Eddington ratios, cooling becomes inefficient, also puffing the disk up. So, the degree to which warps can affect accretion can also be taken as a function of Eddington ratio. 

A particularly interesting laboratory for our accretion mechanisms may be TDE debris disks. This is for a few reasons. First, TDEs have randomly oriented inclinations (and thus an average tilt angle of $\sim60^\circ$). Second, debris disks are formed from material supplied between the stream self-intersection point and the periastron of the tidal disruption, making them much smaller in radial extent than AGN or XRBs \citep{dai_2018,andalman_2022}. Third, although the accretion rate is initially highly super-Eddington, it falls off as $\propto t^{-5/3}$ after its peak, leading to a sharp drop in the scale height \citep{rong_feng_2014,tchekhovskoy_2014,piran_2015}. These considerations suggest that warps may be a primary driver of accretion in TDE debris disks, meriting further study. 

\subsection{Observational implications}
\label{sec:discussion:obs}

The accretion structure and mechanisms that we have described in the preceding sections will necessarily alter the emission of accreting black holes. One of those most interesting effects of tearing and warp-driven accretion is the resulting variability. While the fractional (say, $\sim20-40\%$) broadband variability of accretion disks \citep{gaskell_2004,uttley_2005a,uttley_2005b} can occur between viscous and dynamical timescales \citep[by, for instance, a stochastic dynamo action, e.g.][]{Hogg_2016}, larger structural changes in an accretion disk are typically expected to occur on a viscous timescale. This is, in turn, limited by the value of $\alpha$, which is thought to be $\lesssim0.1-1$ in standard \cite{nt73} disks. However, there is a growing sample of observations of active galactic nuclei (AGN) that exhibit extreme luminosity variations on timescales of months to years, while the viscous timescale can be hundreds to thousands of years. This extreme variability is also not rare; it is exhibited in an estimated $\sim30-50\%$ of quasars \citep{rumbaugh_2018}. These so-called `Changing-look' AGN (CL AGN, \citealt{matt_2003}, see also `quasi-periodic eruptions', \citealt{miniutti_2019}) are difficult to reconcile with theory, leading some to proclaim a `viscosity crisis' in AGN disks \citep{lawrence_2018}. To explain CL AGN, it's thought that there must be some instability that leads to the catastrophic and rapid destruction of the inner accretion flow. Currently proposed theories generally invoke a radiation pressure instability \citep{lightman_eardley_1974} acting on the inner disk \citep{janiuk_czerny_2002,sniegowska_2020,sniegowska_2022a,sniegowska_2022b}, but others have suggested it is the result of a sudden magnetic flux inversion in the disk \citep{scepi_2021}. We argue that the accretion mechanisms presented here may be a natural way of producing CL AGN. Firstly, as demonstrated in Fig. \ref{fig:flowAlpha}, accretion in our simulation happens on timescales that are at least $10-100$ times shorter than the usual viscous timescale. The repeated depletion of the inner sub-disk will also cause a precipitous drop in the luminosity. This supports the tantalizing hypothesis that some CL AGN may be the observational result of the tearing process \citep[see also][]{nixon_king_2012,raj_2021}. We plan to perform a dedicated comparison of GRMHD disk tearing to CL AGN in an upcoming work.     

Quasi-periodic oscillations (QPOs) form another commonly observed, yet poorly understood, class of accretion disk variability. QPOs are variable signals usually observed in the power spectra of X-ray binaries (XRBs) \citep{vanderklis_1985,mucciarelli_2006,gierlinski_2008}, but have also been observed in TDEs and AGN as well \citep{pasham_2019,smith_2021}. The underlying cause of the various kinds of QPOs remains elusive, but possible explanations include the Lense-Thirring precession of tilted disks \citep{stella_1998,stella_1999,fragile_2016} or trapped modes excited by warped or eccentric disks \citep{okazaki_1987,kato_2004,ferreira_ogilvie_2008,ferreira_ogilvie_2009,dewberry_2020a,dewberry_2020b}. A recent work in our collaboration, \cite{gibwa_2022}, has performed a separate analysis on this same simulation and have found evidence of both low-frequency (LF) and high-frequency (HF) QPOs. The HFQPOs were associated with radial epicyclic oscillations of the inner sub-disk (which were not analyzed in this work) and the LFQPOs were associated with geometric effects due to the precession of the inner sub-disk. The variability associated with QPOs are related to, but separate from, any longer-term variability due to the recurrent depletion of the inner sub-disk due to streamers from the tear. 

Another consideration is the emission produced by streamers. Streamers naturally produce hot, low-density features that surround the inner sub-disk (Fig. \ref{fig:PlungingInstantaneous}). This is reminiscent of the usual accretion disk corona and may result in enhanced hard X-ray emission due to the up-scattering of thermal photons emitted by the inner sub-disk. This may help explain the rapid evolution of the X-ray corona observed in some CL AGN \citep[e.g.,][]{ricci_kara_2020} or contribute to the hard emission observed in XRB state transitions \citep{esin_1997,remillard_mcclintock}, which merits a dedicated study of coronal emission during tearing events. 

\subsection{Summary}
We have performed an analysis of a 3D GRMHD simulation of a highly tilted accretion disk around a rapidly rotating black hole, performed at extremely high resolution. We have focused mainly on how the warping and subsequent tearing of the accretion disk impacts its geometry and introduces new dissipation mechanisms that drive rapid accretion. Our main findings in this work are as follows,
\begin{enumerate}
    \item \textbf{Warped accretion disks drive structural oscillations both vertically and radially.} The vertical oscillations manifest as extreme expansions and compressions of the scale height twice an orbit (\secref{sec:geometry:oscillations}). The radial oscillations manifest as eccentric streamlines above and below the midplane of the disk. The argument of periapsis for fluid parcels above and below the disk midplane is out of phase by $\approx180^\circ$. This oscillating solution precesses rigidly with the disk at larger distances ($\gtrsim10-20\,\rg$), but becomes twisted in the rapidly evolving inner sub-disk.
    \item \textbf{The oscillating scale height of the disk results in nozzle shocks twice an orbit, dissipating orbital energy and driving rapid accretion.} Extreme compressions can shock the gas, leading to enhanced dissipation (\secref{sec:accretion:nozzleshocks}). We refer to these as `nozzle shocks' as they are similar to those usually studied in tidal disruption events. These nozzle shocks can lead to rapid accretion, with $\alpha_{\rm eff}\sim10-100$, well in excess of that predicted in standard thin disk models, where it is thought that $\alpha\lesssim0.1-1$.  
    \item \textbf{Disk tearing results in the formation of low angular momentum streamers which rain down on the inner sub-disk and drive further accretion.} When the disk tears, it can lead to the rapid accretion of the inner sub-disk at the tearing radius (\secref{sec:accretion:plunging}). We have attributed this to the cancellation of misaligned angular momentum. This causes low-angular momentum `streamers' to form at the tear and rain down on the inner sub-disk. The addition of low-angular momentum gas to the inner sub-disk causes it shrink to conserve total angular momentum. This can results in high accretion rates, even in the absence of nozzle shocks or a warped inner disk. 
\end{enumerate}
There are several future directions that we would like to consider before concluding. Firstly, the evolution of warped disks is governed by torques acting on both the parallel and perpendicular components of angular momenta, which are mainly determined by the local dissipation mechanisms in the disk. While we have demonstrated that novel dissipation mechanisms (nozzle shocks, tearing and streamers) play an important role in driving the evolution of our disk, we have done so for a single simulation, and it is unknown how these mechanisms depend on parameters such as the initial tilt and thickness of the disk. Insight could be provided by performing more simulations across the relevant parameter space and by the expansion of existing analytic models to include these dissipation mechanisms. A major frontier to explore is also radiation physics. In this work, we use a predefined cooling function, which simplifies the thermodynamics of the disk. This merits the inclusion of dedicated radiation schemes, which we do in L22. Finally, it is ultimately most important to be able to connect our results to observations, which can be accomplished by producing synthetic observations from simulation results such as those presented here.


\begin{acknowledgments}
We thank Y. Lithwick and G. Lodato for useful discussions.  An award of computer time was provided by the Innovative and Novel Computational Impact on Theory and Experiment (INCITE), OLCF Director's Discretionary Allocation, and ASCR Leadership Computing Challenge (ALCC) programs under award PHY129. This research used resources of the Oak Ridge Leadership Computing Facility, which is a DOE Office of Science User Facility supported under Contract DE-AC05-00OR22725. This research used resources of the National Energy Research
Scientific Computing Center, a DOE Office of Science User Facility
supported by the Office of Science of the U.S. Department of Energy
under Contract No. DE-AC02-05CH11231 using NERSC award
ALCC-ERCAP0022634. We acknowledge PRACE for awarding us access to JUWELS Booster at GCS@JSC, Germany. ML is supported by the John Harvard Distinguished Science Fellowship and the ITC fellowship.  NK is supported by an NSF Graduate Research Fellowship. JJ and AT acknowledge support by the NSF AST-2009884 and NASA 80NSSC21K1746 grants.
GM is supported by a Netherlands Research School for Astronomy (NOVA), Virtual Institute of Accretion (VIA) postdoctoral fellowship. AT was supported by the National Science Foundation grants 
AST-2206471, 
AST-2009884, 
AST-2107839, 
AST-1815304,  
OAC-2031997, 
and AST-1911080. 
\end{acknowledgments}

\appendix
\section{Analyzing Misaligned Accretion Disks}
\label{app:misaligned_analysis}
\begin{figure}
    \centering
    \includegraphics[width=0.8\textwidth]{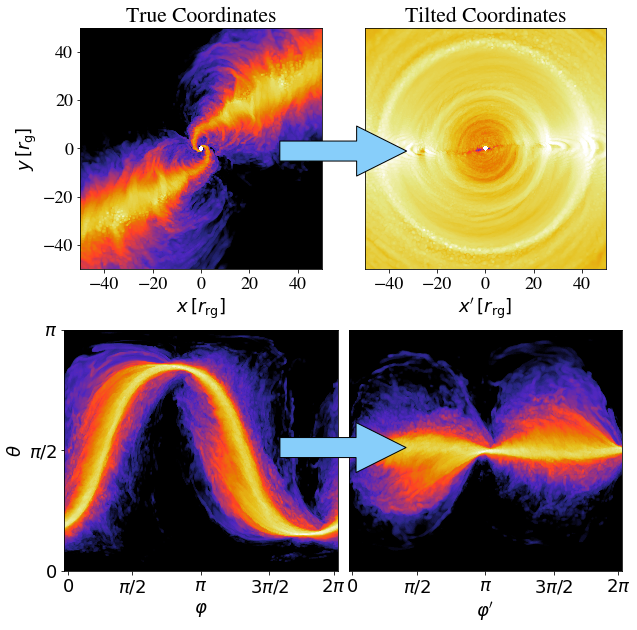}
    \caption{We depict the conversion of global (`true') coordinates to our radially-dependent tilted coordinates. We do so by showing snapshots of $\rho$ in both coordinate systems. The tilted coordinates are indicated by primes and are a function of the local tilt and precession angles (Eq. \ref{eq:app:transformation_matrix}). At all radii in the tilted coordinate system, $\hat{z}^\prime$ is parallel to the local angular momentum vector, $\varphi^\prime=0$ is set to the local precession angle, and $\theta^\prime=\pi/2$ is set to the local midplane of the disk.}
    \label{fig:app:tilt_algo}
\end{figure}

The complicated geometry of warped accretion disks can make them difficult to visualize and analyze. To alleviate this, we perform conversions from `true' ($r,\theta,\varphi$) to `tilted' ($r,\theta^\prime,\varphi^\prime$) coordinates. There are two parts to this conversion. First, we have to convert the grid itself. To do this, we calculate radial profiles of the local angular momentum vector and use its orientation to determine the corresponding tilt and precession angles, following \cite{fragile_anninos_2005}. Then, we rotate a $\theta-\varphi$ grid at each radius such that the precession angle is at $\varphi^\prime=0$ and the tilt angle is $=0^\circ$ (i.e., the disk midplane is at $\theta^\prime=\pi/2$). We then interpolate all relevant quantities to the rotated grid. This is visualized for the fluid frame gas density $\rho$ in Fig. \ref{fig:app:tilt_algo}. In the top row, we can see that an $x-y$ slice in true coordinates becomes a face-down image of the disk in tilted coordinates. In the bottom row, we see that a $\theta-\varphi$ slice in true coordinates shows an annulus rocking up and down due to its tilt angle, while in tilted coordinates the annulus is flattened. 

The second part of this conversion is the coordinate transformation of four-vectors from true to tilted coordinates, which for instance reads $u^{\mu^\prime} = \frac{\partial \chi^{\mu^\prime}}{\partial \chi^\mu}u^\mu$ for the four-velocity. The specific matrix elements of the transformation are, 

\begin{equation} \frac{\partial \chi^{\mu^\prime}}{\partial \chi^\mu} = \begin{bmatrix}
1 & 0 & 0 & 0 \\
0 & {\rm cos}(\mathcal{T}){\rm cos}(\mathcal{P}) & -{\rm cos}(\mathcal{T}){\rm sin}(\mathcal{P}) & -{\rm sin}(\mathcal{T}) \\
0 & {\rm sin}(\mathcal{P}) & {\rm cos}(\mathcal{P}) & 0 \\
0 & {\rm sin}(\mathcal{T}){\rm cos}(\mathcal{P}) & -{\rm sin}(\mathcal{T}){\rm sin}(\mathcal{P}) & {\rm cos}(\mathcal{T}) 
\end{bmatrix}
\label{eq:app:transformation_matrix}
\end{equation}
where $\mathcal{T}$ and $\mathcal{P}$ are the tilt and precession angles, respectively.
This is a purely Newtonian rotation that leaves the time component of our four-vectors unchanged. In tilted coordinates, the azimuthal velocity $u^{\varphi^\prime}$ is aligned with the rotation of its annulus in an average sense. Correspondingly, the vertical velocity $u^{\theta^\prime}$ averaged over an annulus is approximately zero. 



\bibliographystyle{aasjournal}
\bibliography{references}

\end{document}